\documentclass[prl,aps,floatfix,twocolumn]{revtex4}
\usepackage{epsfig}
\begin{document}

\title{
Theory of spin-polarized transport in semiconductor heterojunctions:
Proposal for spin injection and detection in silicon
}

\author{Igor \v{Z}uti\'{c}$^{1,2}$, Jaroslav 
Fabian$^{3}$\footnote{Present address: Institute for Theoretical Physics, 
University of Regensburg, 93040 Regensburg, Germany},
and Steven C. Erwin$^1$}
\affiliation{$^1$Center for Computational Materials 
Science, Naval Research Laboratory, Washington, D.C. 20375 \\ 
$^2$Condensed Matter Theory Center, University of Maryland, College Park, 
Maryland 20742 \\
$^3$Institute for Theoretical Physics, 
Karl-Franzens University, Universit\"atsplatz 5, 8010 Graz, Austria}

\begin{abstract}
Spin injection and detection in silicon is a difficult problem, in
part because the weak spin-orbit coupling and indirect gap preclude
using standard optical techniques.  We propose two ways to overcome
this difficulty, and illustrate their operation by developing a model
for spin-polarized transport across a heterojunction. We find that
equilibrium spin polarization of holes leads to a strong modification
of the spin and charge dynamics of electrons, and we show how the
symmetry properties of the charge current can be exploited to detect
spin injection in silicon using currently available techniques.
\end{abstract}
\vspace{-0.6cm}
%]
%\newpage
\maketitle
In addition to its central role in conventional electronics, silicon
has spin-dependent properties (such as long spin relaxation and
decoherence times) that could be particularly
useful in spin-based quantum-information processing and 
spintronics~\cite{Zutic2004:RMP}. Unfortunately, the underlying origins 
of these attractive properties---the indirect band gap, weak spin-orbit
coupling, and extremely small concentration of paramagnetic 
impurities~\cite{Tyryshkin2003:PRB}---also preclude using the standard 
optical methods of spin injection
and detection in semiconductors.

Circularly polarized light can be used to polarize carriers in
semiconductors with a direct band gap.  Moreover,
both the direction and the magnitude of optically generated 
charge currents~\cite{Ganichev2002:PRL,Ganichev2002:N} and
pure spin currents~\cite{Stevens2003:PRL,Hubner2003:PRL} 
can be controlled optically.
In the reverse process, the presence of polarized carriers in a
direct-gap semiconductor can be detected by 
measuring the circular polarization of the recombination
light~\cite{Meier:1984,Zutic2004:RMP}.  
Typical detection schemes use spin light-emitting diodes (LED's),
where the selection rules for radiative recombination processes 
can be used to relate the circular polarization of the emitted light to
the spin polarization of the 
carriers~\cite{Fiederling1999:N,Jonker2000:PRB,Young2002:APL,Jiang2003:PRL}.

\begin{figure}
\centerline{\psfig{file=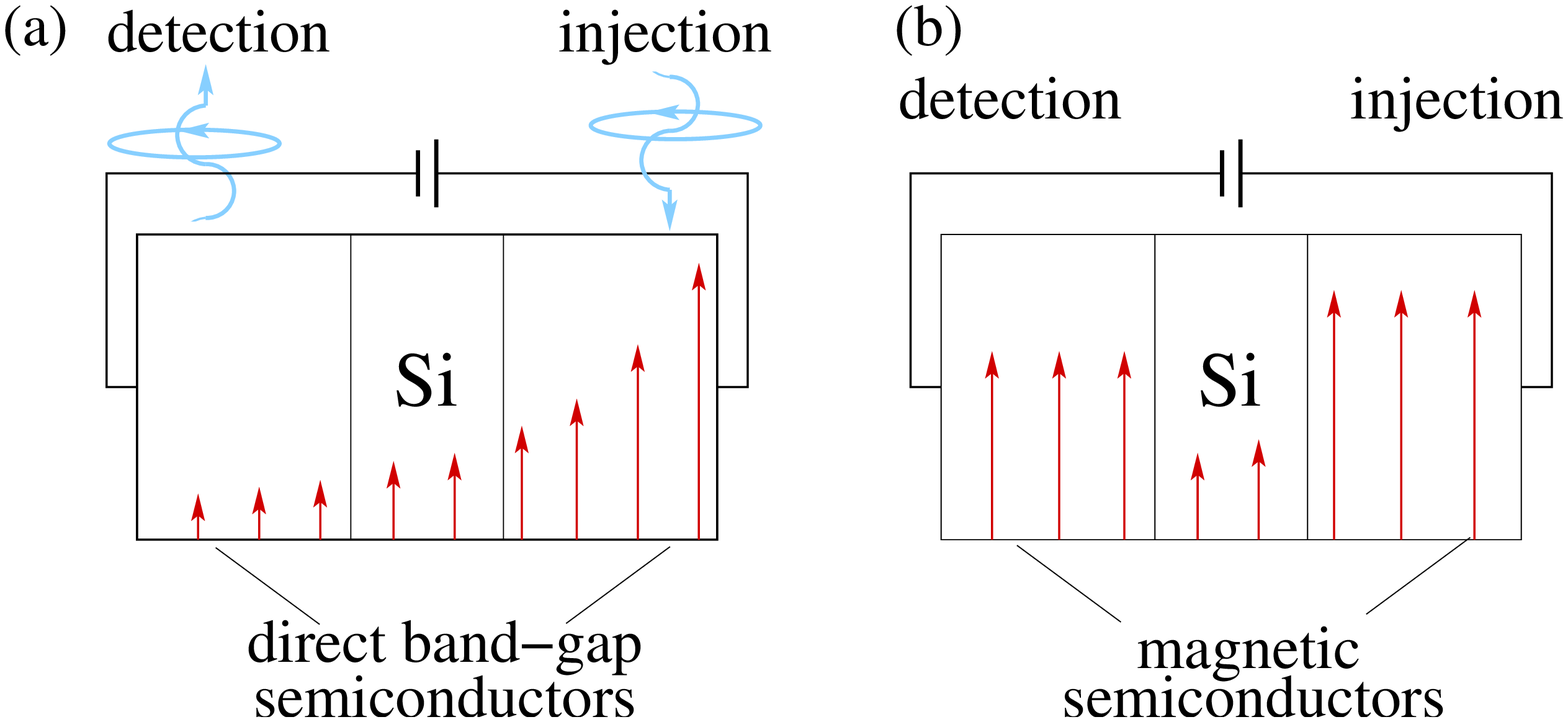,width=1\linewidth,angle=0}}  
\caption{Proposed schemes for spin injection and detection in silicon.
(a) Optical realization based on radiative processes (excitation
for spin injection and recombination for detection) in direct-gap
semiconductors surrounding silicon. Arrows depict the spatial decay of
nonequilibrium spin. (b) Electrical realization based on spin
splitting and net spin density (magnetization) in the two magnetic
regions. The relative orientation of the nonequilibrium spin in Si and
the equilibrium spin in the magnetic regions influences the magnitude of a
charge current or an open-circuit voltage.  Other realizations are
also possible by combining schemes (a) and (b).}
\label{fig:1}
\end{figure}

For silicon, the indirect band gap makes direct application of these
techniques problematic.  We propose two approaches to overcome this
difficulty.  The first, shown in Fig.~\ref{fig:1}(a), is based on
high-quality heterojunctions between Si and a direct-gap
semiconductor.  In such a heterojunction, optical techniques could be
readily employed in the direct-gap semiconductor to circumvent the
problems with spin injection and detection in Si.  The key
prerequisite for such a proposal is an interface with Si that would
not be detrimental to the spin transport. This is a nontrivial
undertaking, as the lattice mismatch between Si and most direct-gap
semiconductors typically leads to low-quality interfaces with a high
density of interfacial defects.  Nevertheless, there has been recent
progress in fabricating high-quality GaAs/Si interfaces (despite the
4$\%$ lattice mismatch)~\cite{Yonezu2002:SST}, and GaP$_{1-x}$N$_x$
leads to even smaller mismatch (below $1\%$)~\cite{Vurgaftman2001:JAP}. 
Recent studies of charge
transport in GaAs/Si heterojunctions 
suggest the feasibility of the scheme shown in 
Fig.~\ref{fig:1}(a). In particular, GaAs/Si heterojunctions displayed
$I$-$V$ characteristics of an ideal diode~\cite{Aperathitis1996:ASS}, and
optical excitations were studied in GaAs LED's grown 
on Si~\cite{Taylor2001:JAP}.

In the second approach, shown in Fig.~\ref{fig:1}(b), magnetic
semiconductors approximately lattice matched with Si could be used for
spin injection and detection \cite{gasi}. For example, 
the Mn-doped chalcopyrite
ZnGeP$_2$ (mismatch $< 2\%$)~\cite{Ishida2003:PRL,Cho2002:PRL}
has been reported to be ferromagnetic at
room temperature. Another Mn-doped chalcopyrite, ZnSiP$_2$, was
recently predicted~\cite{Erwin2004:NM} to be ferromagnetic, as well as
highly spin polarized and closely lattice-matched with Si (mismatch
$<1\%$). Mn doping of the chalcopyrite alloy ZnGe$_{1-x}$Si$_x$P$_2$
would likely lead to an exact lattice match, since the lattice
constant of Si is between those of closely matched ZnSiP$_2$ and
ZnGeP$_2$.
 
\begin{figure}
\centerline{\psfig{file=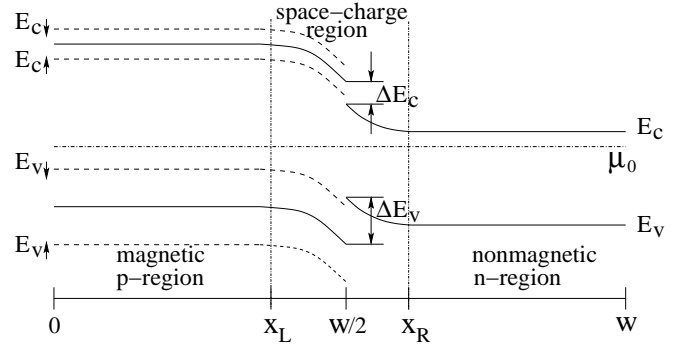,width=1\linewidth,angle=0}}  
\caption{Band diagram for a magnetic heterojunction. 
In equilibrium, the chemical potential $\mu_0$ is constant.
Conduction- and valence-band edges ($E_c$ and $E_v$) are spin-split in
the magnetic $p$-region, while there is no spin splitting in the
nonmagnetic $n$-region (corresponding to Si).  For a sharp doping
profile, there are generally discontinuities ($\Delta E_c$ and $\Delta
E_v$) in the conduction
and valence bands at $x=w/2$.}
\label{fig:2}
\end{figure}

To capture the main features of spin-polarized transport across a
heterojunction we formulate here a model, representing both schemes
in Fig.~\ref{fig:1}, that can be solved analytically.  One can model
the right (injecting) electrode by appropriate boundary conditions,
and hence we
focus on the two left regions that define the heterojunction in our
model.  A heterojunction is doped with ionized acceptors and donors of
density $N_a$ and $N_d$. Codoping with magnetic
impurities would additionally introduce a net spin, but need
not change the number of carriers.  An inhomogeneous distribution of
$N_{a,d}$ implies a large deviation from local charge neutrality,
so that Poisson's equation must be explicitly solved.  For nondegenerate
doping levels (Boltzmann statistics) the spin-resolved
quasiequilibrium electron and hole densities are
\begin{equation}
n_\lambda=\frac{N_c}{2} e^{-[E_{c\lambda}-\mu_{n\lambda}]/k_BT}, \quad
p_\lambda=\frac{N_v}{2} e^{-[\mu_{p\lambda}-E_{v\lambda}]/k_BT},
\label{eq:ncv}
\end{equation}
where $\lambda$ = $+1$ for spin up ($\uparrow$) and
$-1$ for spin down ($\downarrow$). The total electron density
$n=n_\uparrow+n_\downarrow$ 
can also be decomposed as a sum of equilibrium 
and nonequilibrium parts, $n=n_0+\delta n$. We define 
the electron spin density $s_n=n_\uparrow-n_\downarrow$ 
and the spin polarization $P_n=s_n/n$, with an analogous notation
for holes. 
In Eq.~(\ref{eq:ncv}), subscripts $c$ and $v$
label quantities pertaining to conduction and valence bands.
The corresponding effective density of states are 
$N_{c,v}=2(2 \pi m_{c,v} k_B T/h^2)^{3/2}$, where
$m_{c,v}$ are effective masses.
The spin-$\lambda$ conduction band edge (see Fig.~\ref{fig:2}) 
$E_{c\lambda}=E_{c0}-q\phi-\lambda q \zeta_c$ differs from its
nonmagnetic bulk value $E_{c0}$ because of the electrostatic
potential $\phi$ and the spin splitting $2 q \zeta_c$, which
parameterizes Zeeman or exchange splitting due to magnetic impurities 
and/or an applied magnetic field~\cite{Zutic2004:RMP}.
Here, $\Delta E_c$ 
is the conduction band edge discontinuity and
$\mu_{n \lambda}=\mu_0+\delta \mu_{n
\lambda}$ is the chemical potential for spin-$\lambda$ electrons.
An analogous notation holds for the valence band and holes.

We assume transport across the interface is dominated by
drift-diffusion, so that the spin-resolved charge-current densities are
\begin{eqnarray}
\label{eq:jn}
{\bf J}_{n\lambda}&=
&q\bar{\mu}_{n\lambda} n_{\lambda} \nabla E_{c \lambda}
+qD_{n\lambda} N_c \nabla (n_\lambda/N_c),
\\ \label{eq:jp}
{\bf J}_{p\lambda}&=&
q\bar{\mu}_{p\lambda} p_{\lambda} \nabla E_{v \lambda}
-qD_{p\lambda} N_v \nabla (p_\lambda/N_v),
\end{eqnarray}
where $\bar{\mu}$ and $D$ are mobility and diffusion
coefficients.
We note that ``drift terms'' have quasi-electric fields 
$\propto \nabla E_{c,v \lambda}$ that are generally
spin-dependent ($\nabla \zeta_{c,v} \neq 0$ is referred to as
a magnetic drift~\cite{Zutic2002:PRL})
and different for conduction and valence bands.
In contrast to  homojunctions, additional
``diffusive terms'' arise due to the spatial dependence
of $m_{c,v}$, and therefore of $N_{c,v}$. 

We write 
the continuity equation for $n_\lambda$ as
\begin{eqnarray}
-\partial n_\lambda/\partial t&+&
\nabla\cdot {\bf J}_{n\lambda}/q=+ r_{\lambda}(n_{\lambda}
p_\lambda -n_{\lambda 0} p_{\lambda 0}) \\ \nonumber
&+& [n_{\lambda}-n_{-{\lambda}}-\lambda P_{n0} n]/2\tau_{sn}
-G_{\lambda},
\label{eq:ncont}
\end{eqnarray}
with an analogous equation for $p_\lambda$.
Here, $r_\lambda$ is the recombination rate of
spin-$\lambda$ carriers;
$\tau_{s n,p}$ is spin relaxation
time for electrons and holes; and $G_\lambda$ is the 
photoexcitation rate due to 
electron-hole pair generation and optical orientation
(when $G_\uparrow \neq G_\downarrow$)~\cite{Zutic2004:RMP,Meier:1984}.
Spin relaxation equilibrates carrier spin while preserving nonequilibrium 
carrier density~\cite{Meier:1984}, so that for nondegenerate
semiconductors we have $P_{n0}=\tanh(q \zeta_c/k_BT)$.

We make several assumptions allowing us to solve this model
analytically.
We focus here on the steady-state low-injection regime, at applied bias
$|V|< min (|E_{c\lambda}-E_{v\lambda}|)$.
Spin-orbit coupling in the valence band typically leads to a much faster spin 
relaxation of holes than electrons (3-4 orders faster in 
GaAs~\cite{Zutic2004:RMP}), and so it is reasonable to consider that 
the spin of holes is in equilibrium ($\delta s_p=0$ and 
$P_p=P_{p0}$). As a result,
only the nonequilibrium electron
spin density ($\delta s_n \rightarrow \delta s$) and the minority carrier density 
need to be calculated throughout the heterojunction.
We assume a sharp doping 
profile $N_d(x)-N_a(x)$; referring to Fig.~\ref{fig:2}, this leads to
a discontinuous change in materials parameters at $x=w/2$. We take
$N_{c,v}$, $\bar{\mu}$, $D$, and the permittivity $\epsilon$ to be
constant outside the space-charge region $x_L<x<x_R$, and hence label them
by indices $L$ and $R$. 
The width of a space-charge region is $x_R-x_L \propto (V_{bi}-V)^{1/2}$,
where the built-in voltage is 
$q V_{bi}=-\Delta E_c+k_B T \ln (n_{0R} N_{cR}/n_{0L} N_{cL})$; we
note that the discontinuity
$\Delta E_{c,v}$ can be accurately measured at 
interfaces with Si~\cite{Marka2003:PRB}. 
Equations 
(\ref{eq:jn}) and (\ref{eq:jp}), together with the continuity 
equations, reduce to diffusion-like equations 
for $\delta n, \delta s$ in the $p$-region and $\delta p, \delta s$ 
in the $n$-region. 
For the the (magnetic) $p$-region, 
we find that the
spatial dependence of both $\delta n$ and $\delta s$
are described by two distinct decay lengths; this is in marked contrast
to previously studied cases~\cite{Zutic2004:RMP}.
These decay lengths can be written as
\begin{eqnarray}
\label{eq:kappa}
& &\kappa^{-1}, \chi^{-1} =
[(L_\uparrow^{-2}+L_\downarrow^{-2}+L_s^{-2})/2 \\ 
&\pm&
[(L_\uparrow^{-2}-L_\downarrow^{-2}-P_{n0} L_s^{-2})^2
+(1-P_{n0}^2)L_s^{-4}]^{1/2}/2]^{-1/2},   \nonumber
\end{eqnarray}
where the upper  
sign refers to $\kappa^{-1}$; 
$L_\lambda=(D_n/r_\lambda p_{\lambda0})^{1/2}$ are the
electron diffusion lengths; and $L_s=(D_n\tau_{sn})^{1/2}$ 
is the electron spin diffusion length.

\begin{figure}
\centerline{\psfig{file=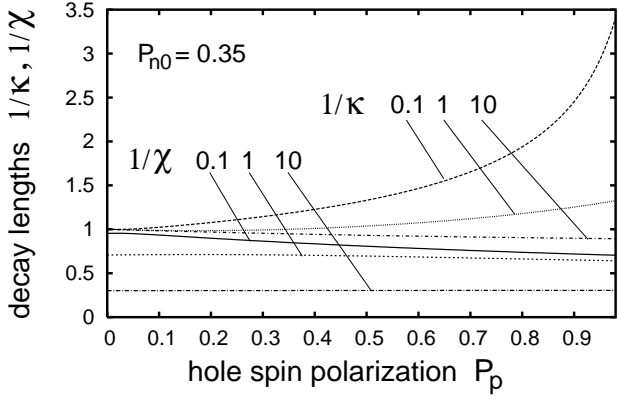,width=1\linewidth,angle=0}}  
\caption{Decay lengths $\kappa^{-1}$ and $\chi^{-1}$ in the 
magnetic $p$-region, 
normalized to the electron diffusion length $L_n=(D_n \tau_n)^{1/2}$,
as a function of hole-spin polarization for fixed electron-spin
polarization $P_{n0}=0.35$. Each curve is labeled by the  
ratio of electron lifetime to spin-relaxation time, $\tau_n/\tau_{sn}$.}
\label{fig:3}
\end{figure}

It is instructive to consider the regime of 
spin-unpolarized holes, appropriate
for
the scheme shown in Fig.~\ref{fig:1}(a). In this regime we have
$p_{\uparrow 0}=p_{\downarrow 0}=N_a/2$ and 
$r_\uparrow=r_\downarrow=r/2$ (in a nondegenerate regime 
$r_\uparrow \approx r_\downarrow$ even for 
$P_p \neq 0$~\cite{Lebedeva2003:JAP}). 
It follows from Eq.~(\ref{eq:kappa}) that $\kappa^{-1}$ reduces to
$L_{sn}=(D_n T_s)^{1/2}$, where $L_{sn}$ is the effective
electron-spin diffusion length and $T_s=(r N_a + 1/\tau_{sn})^{-1}$ is
the electron-spin lifetime. Analogously, $\chi^{-1}$ reduces to the
electron diffusion length $L_n=(D_n \tau_n)^{1/2}$, where $\tau_n=1/r
N_a$ is the electron lifetime~\cite{Meier:1984}.  Thus, in this regime
$\kappa^{-1}$ and $\chi^{-1}$ separately determine
the decay lengths for $\delta s$ and $\delta n$, respectively. 
For the more general case, in
Fig.~\ref{fig:3} we show $\kappa^{-1}$ and $\chi^{-1}$ as function of
$P_p$. From the 
behavior of the decay lengths for $\delta n$ and $\delta s$, we 
conclude that polarization of the equilibrium hole spins leads to a
strong modification of charge and spin dynamics of electrons.

We turn now to the more general case of spin detection using magnetic
semiconductors, shown in Fig.~\ref{fig:1} (b), and solve the
corresponding problem of spin-polarized transport across the
heterojunction in Fig.~\ref{fig:2}. We impose the ohmic boundary
conditions $\delta n = \delta s=0$ at $x=0$, and include optical or
electrical carrier and spin injection through the boundary conditions
$\delta p \neq 0$ and $\delta s \neq 0$ at $x=w$.
To match the chemical potentials $\mu_{n,p\lambda}$
at $x_L$ and $x_R$ (which is an accurate approximation of a full
numerical solution in magnetic {\it p-n} 
junctions~\cite{Zutic2002:PRL}) requires satisfying the
self-consistency condition
$P^L_n=(P^L_{n0}+\delta P^R_{n})/
(1+P^L_{n0} \delta P^R_{n})$, where $\delta P_n^R = \delta s_R/N_d$
is determined from the continuity of the spin current,
$D_{nL} d (\delta s_L)/dx=D_{nR} d(\delta s_R)/dx$. 
Consistent with this matching,  
a generalization
of Shockley's relation~\cite{Fabian2002:PRB} is
$\delta n_{L}=n_{0L}[\exp(qV/k_BT)-1]+s_{0L}\exp(qV/k_BT) \delta P^R_{n}$
and
$\delta s_{L}=s_{0L}[\exp(qV/k_BT)-1]+n_{0L}\exp(qV/k_BT) \delta P^R_{n}$,
such that $\delta n_{L}, \delta s_{L}$ and 
$\delta n_{R}, \delta s_{R}$ can be considered as boundary conditions
in the $p$- and $n$-region, respectively. 
For the $p$-region we then obtain
\begin{eqnarray}
\label{eq:density}
\delta n, \delta s =&-&C_{\kappa n,s} \frac{(\chi^2+a) \delta n_ L 
+ b \delta s_L} {(\kappa^2-\chi^2) \sinh (\kappa x_L)} \sinh (\kappa x) \\
&+&C_{\chi n,s} \frac{(\kappa^2+a)  \delta n_L + b \delta s_L}
{ (\kappa^2-\chi^2) \sinh (\chi x_L)} \sinh (\chi x), \nonumber
\end{eqnarray}
where $C_{\kappa n}=C_{\chi n}=1$ for $\delta n$;   
$C_{\kappa s}=-(\kappa^2+a)/b$ and $C_{\chi s}=-(\chi^2+a)/b$ for
$\delta s$; and we have defined
$a=-(L_\uparrow^{-2}+L_\downarrow^{-2})/2$ and 
$b=-(L_\uparrow^{-2}-L_\downarrow^{-2})/2$.

In the nonmagnetic $n$-region (representing Si), the total charge
current $J$ is the sum of
minority carrier currents at $x_L$ and $x_R$, $J=J_{nL}+J_{pR}$, in
analogy to Shockley's
formulation~\cite{Shockley:1950}.
Here, $J_{nL}=q D_{nL} d (\delta n_L)/dx$, and 
$d (\delta n_L)/dx$ can be evaluated using Eq.~(\ref{eq:density}).
A straightforward method for detecting injected spin in Si 
follows from the symmetry
properties of the different contributions to the charge current under
magnetization reversal.
By reversing the equilibrium spin polarization using
a modest external magnetic field ($P_{n0},P_p \rightarrow -P_{n0},-P_p$)
it follows from Eq.~(\ref{eq:kappa}) that 
$\kappa,\chi \rightarrow \kappa,\chi$ and $a,b  \rightarrow a,-b$.
A part of $J_{nL}$, odd under such reversal, can be identified
as the spin-voltaic current~\cite{Zutic2002:PRL},
\begin{equation}
J_{sv} \propto q D_{nL} n_{0L}\exp(qV/k_BT) \delta P^R_{n}
/(\kappa^2-\chi^2),
\label{eq:sv}
\end{equation}
which originates from the interplay of the equilibrium 
and nonequilibrium (injected) spin polarization in the
$p$- and the $n$-region, respectively. 
Measurements of  $J(V,P_{n0},P_p)-J(V,-P_{n0},-P_p)
=2 J_{sv}(V,P_{n0},P_p)$ would then provide: (1) cancellation
of contributions to the charge current that are not related
to the injected spin in Si; (2) a choice of 
$V$ to facilitate
a sufficiently large $J_{sv}$ for accurate detection. 
Alternatively, for the optical detection scheme of Fig.~\ref{fig:1}(a),
one would consider the limit $P_{n0}=P_p=0$ 
in Eq.~(\ref{eq:density}), 
and evaluate $P_n(x)$.
 
\begin{figure}
\centerline{\psfig{file=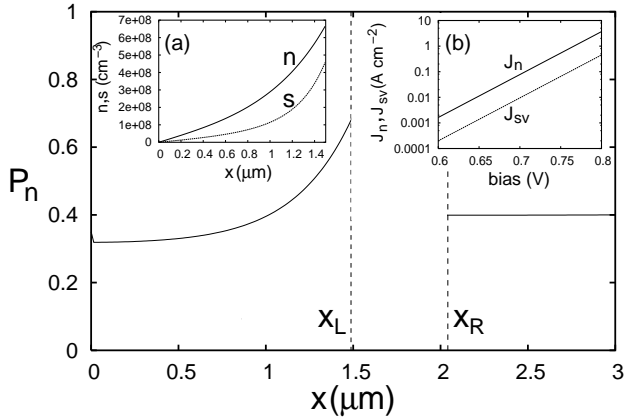,width=1\linewidth,angle=0}}  
\caption{Electron-spin polarization, $P_n$, for a 
magnetic heterojunction at forward bias $V$=0.7 V.
The equilibrium spin polarizations in 
the $p$-region 
are $P_{n0}=0.35$ and $P_p=0.8$, while the injected spin polarization
in  
the $n$-region is $\delta P_n=0.4$ at $w=3$ $\mu$m. Inset (a):
profile of the electron carrier and spin density in the $p$-region.
Inset (b): voltage dependence of the electron charge current
$J_n$ and its part due to the nonequilibrium spin (the spin-voltaic 
current $J_{sv}$).}
\label{fig:4}
\end{figure}

We illustrate the effects of spin injection and detection 
in Si using the heterojunction in Fig.~\ref{fig:2}, 
and show the results in Fig.~\ref{fig:4}.
We use a standard set of parameters for Si doped with 
$N_d=10^{17}$ cm$^{-3}$~\cite{Levinshtein:1996}: 
$N_{cR}=3.2\times 10^{19}$ cm$^{-3}$, $D_{nR}=4 D_{pR}=20$ cm$^2/s$,
$\epsilon_R=11.7$, $\tau_p=10^{-7}$ s, 
an intrinsic carrier density~\cite{Fabian2002:PRB} 
$n_{iR}=10^{10}$ cm$^{-3}$, and estimated spin relaxation time in Si
as $\tau_{sR}=10^{-7}$ s. Circularly polarized light in
zinc-blende semiconductors, such as GaAs, generates spin polarization
of up to $\delta P_n=0.5$, which can be increased even further 
using strain or quantum confinement~\cite{Zutic2004:RMP,Meier:1984}. 
We model the effects of spin injection from a neighboring direct-gap
semiconductor, as depicted in Fig.~\ref{fig:1}, 
by assuming $\delta P_n=0.4$ at $x=w$, and we set $\Delta E_c =0.2$.
For a ferromagnetic semiconductor in the
$p$-region we choose $N_a=10^{15}$ cm$^{-3}$, $N_{cL}=10^{19}$ cm$^{-3}$, 
$D_{nL}=5$ cm$^2/s$, $\epsilon_L=10$, $\tau_n=10^{-9}$ s, 
$n_{iL}=10^{6}$ cm$^{-3}$, and $\tau_{sL}=10^{-10}$ s. 
We parameterize the spin splitting of carrier bands
(see Fig.~\ref{fig:2}) with $P_{n0}=0.35$ and $P_p=0.8$
which, for Boltzmann statistics, corresponds to a splitting
of $q \zeta_v \approx 3 q \zeta_c \approx k_BT$~\cite{temp}.
The effect of decay lengths in the $p$-region
($\kappa^{-1} \approx 0.22$ $\mu$m and $\chi^{-1} \approx 0.64$ $\mu$m)
can be seen in the Fig.~\ref{fig:4}(a). In the $n$-region,
$L_{sR}=(D_{nR} \tau_{sR})^{1/2} \approx 14$ $\mu$m implies
a weak spin decay ($\delta P_n^w \approx \delta P_n^R$),
and the self-consistency condition gives an estimate 
$P_n^L \approx (0.35+0.4)/(1+0.35 \times 0.4) \approx 0.66$. 
We find a large ``signal-to-background ratio,'' $J_{sv}/J_n$,
over a range of bias voltages [see Fig.~\ref{fig:4}(b)], and
thus we predict that
the magnitude of this spin-voltaic current---a fingerprint for spin injection and
detection in Si---could readily be detected using
existing experimental techniques~\cite{Jonker2004:PC}.

Our theoretical framework could be also used to calculate the effects
of nonequilibrium spin in other magnetic heterojunctions, 
including Si-based spin transistors~\cite{Thompson2004:P}. 
An extension of our results to the regime of large forward bias
could provide an alternative method to detect nonequilibrium spin 
in Si. Even without a source of spin injection, a nonequilibrium
spin could be generated through the process of spin 
extraction~\cite{Zutic2002:PRL,Bratkovsky2004:JAP} into a neighboring 
magnetic region.
Our findings should also provide useful test cases for developing
more general numerical methods for treating spin-polarized 
transport~\cite{Saikin2003:JAP}.

This work was supported by ONR, DARPA, NSF, and
CNMS ORNL. I. \v{Z}.
acknowledges financial support from the National Research Council.


\vspace{-0.4cm}

\begin{thebibliography}{33}
\expandafter\ifx\csname natexlab\endcsname\relax\def\natexlab#1{#1}\fi
\expandafter\ifx\csname bibnamefont\endcsname\relax
  \def\bibnamefont#1{#1}\fi
\expandafter\ifx\csname bibfnamefont\endcsname\relax
  \def\bibfnamefont#1{#1}\fi
\expandafter\ifx\csname citenamefont\endcsname\relax
  \def\citenamefont#1{#1}\fi
\expandafter\ifx\csname url\endcsname\relax
  \def\url#1{\texttt{#1}}\fi
\expandafter\ifx\csname urlprefix\endcsname\relax\def\urlprefix{URL }\fi
\providecommand{\bibinfo}[2]{#2}
\providecommand{\eprint}[2][]{\url{#2}}

\bibitem[{\citenamefont{{\v{Z}uti\'{c}}
  et~al.}(2004)\citenamefont{{\v{Z}uti\'{c}}, Fabian, and {Das
  Sarma}}}]{Zutic2004:RMP}
\bibinfo{author}{\bibfnamefont{I.}~\bibnamefont{{\v{Z}uti\'{c}}}},
  \bibinfo{author}{\bibfnamefont{J.}~\bibnamefont{Fabian}}, \bibnamefont{and}
  \bibinfo{author}{\bibfnamefont{S.}~\bibnamefont{{Das Sarma}}},
  \bibinfo{journal}{Rev. Mod. Phys.} \textbf{\bibinfo{volume}{76}},
  \bibinfo{pages}{323} (\bibinfo{year}{2004}).

\bibitem[{\citenamefont{Tyryshkin et~al.}(2003)\citenamefont{Tyryshkin, Lyon,
  Astashkin, and Raitsimring}}]{Tyryshkin2003:PRB}
\bibinfo{author}{\bibfnamefont{A.~M.} \bibnamefont{Tyryshkin}},
  \bibinfo{author}{\bibfnamefont{S.~A.} \bibnamefont{Lyon}},
  \bibinfo{author}{\bibfnamefont{A.~V.} \bibnamefont{Astashkin}},
  \bibnamefont{and} \bibinfo{author}{\bibfnamefont{A.~M.}
  \bibnamefont{Raitsimring}}, \bibinfo{journal}{Phys. Rev. B}
  \textbf{\bibinfo{volume}{68}}, \bibinfo{pages}{193207}
  (\bibinfo{year}{2003}).

\bibitem[{\citenamefont{Ganichev
  et~al.}(2002{\natexlab{a}})\citenamefont{Ganichev, Danilov, Belkov, Ivchenko,
  Bichler, Wegscheider, Weiss, and Prettl}}]{Ganichev2002:PRL}
\bibinfo{author}{\bibfnamefont{S.~D.} \bibnamefont{Ganichev}},
  \bibinfo{author}{\bibfnamefont{S.~N.} \bibnamefont{Danilov}},
  \bibinfo{author}{\bibfnamefont{V.~V.} \bibnamefont{Belkov}},
  \bibinfo{author}{\bibfnamefont{E.~L.} \bibnamefont{Ivchenko}},
  \bibinfo{author}{\bibfnamefont{M.}~\bibnamefont{Bichler}},
  \bibinfo{author}{\bibfnamefont{W.}~\bibnamefont{Wegscheider}},
  \bibinfo{author}{\bibfnamefont{D.}~\bibnamefont{Weiss}}, \bibnamefont{and}
  \bibinfo{author}{\bibfnamefont{W.}~\bibnamefont{Prettl}},
  \bibinfo{journal}{Phys. Rev. Lett.} \textbf{\bibinfo{volume}{88}},
  \bibinfo{pages}{057401} (\bibinfo{year}{2002}{\natexlab{a}}).

\bibitem[{\citenamefont{Ganichev
  et~al.}(2002{\natexlab{b}})\citenamefont{Ganichev, Ivchenko, Belkov,
  Tarasenko, Sollinger, Weiss, Wegscheider, and Prettl}}]{Ganichev2002:N}
\bibinfo{author}{\bibfnamefont{S.~D.} \bibnamefont{Ganichev}},
  \bibinfo{author}{\bibfnamefont{E.~L.} \bibnamefont{Ivchenko}},
  \bibinfo{author}{\bibfnamefont{V.~V.} \bibnamefont{Belkov}},
  \bibinfo{author}{\bibfnamefont{S.~A.} \bibnamefont{Tarasenko}},
  \bibinfo{author}{\bibfnamefont{M.}~\bibnamefont{Sollinger}},
  \bibinfo{author}{\bibfnamefont{D.}~\bibnamefont{Weiss}},
  \bibinfo{author}{\bibfnamefont{W.}~\bibnamefont{Wegscheider}},
  \bibnamefont{and} \bibinfo{author}{\bibfnamefont{W.}~\bibnamefont{Prettl}},
  \bibinfo{journal}{{\sl Nature}} \textbf{\bibinfo{volume}{417}},
  \bibinfo{pages}{153} (\bibinfo{year}{2002}{\natexlab{b}}).

\bibitem[{\citenamefont{Stevens et~al.}(2003)\citenamefont{Stevens, Smirl, {R.
  Bhat}, Najmaie, Sipe, and {van Driel}}}]{Stevens2003:PRL}
\bibinfo{author}{\bibfnamefont{M.~J.} \bibnamefont{Stevens}},
  \bibinfo{author}{\bibfnamefont{A.~L.} \bibnamefont{Smirl}},
  \bibinfo{author}{\bibfnamefont{R.~D.} \bibnamefont{{R. Bhat}}},
  \bibinfo{author}{\bibfnamefont{A.}~\bibnamefont{Najmaie}},
  \bibinfo{author}{\bibfnamefont{J.~E.} \bibnamefont{Sipe}}, \bibnamefont{and}
  \bibinfo{author}{\bibfnamefont{H.~M.} \bibnamefont{{van Driel}}},
  \bibinfo{journal}{Phys. Rev. Lett.} \textbf{\bibinfo{volume}{90}},
  \bibinfo{pages}{136603} (\bibinfo{year}{2003}).

\bibitem[{\citenamefont{{H\"{u}bner} et~al.}(2003)\citenamefont{{H\"{u}bner},
  {R\"{u}hle}, Klude, Hommel, {R. Bhat}, Sipe, and {van
  Driel}}}]{Hubner2003:PRL}
\bibinfo{author}{\bibfnamefont{J.}~\bibnamefont{{H\"{u}bner}}},
  \bibinfo{author}{\bibfnamefont{W.~W.} \bibnamefont{{R\"{u}hle}}},
  \bibinfo{author}{\bibfnamefont{M.}~\bibnamefont{Klude}},
  \bibinfo{author}{\bibfnamefont{D.}~\bibnamefont{Hommel}},
  \bibinfo{author}{\bibfnamefont{R.~D.} \bibnamefont{{R. Bhat}}},
  \bibinfo{author}{\bibfnamefont{J.~E.} \bibnamefont{Sipe}}, \bibnamefont{and}
  \bibinfo{author}{\bibfnamefont{H.~M.} \bibnamefont{{van Driel}}},
  \bibinfo{journal}{Phys. Rev. Lett.} \textbf{\bibinfo{volume}{90}},
  \bibinfo{pages}{216601} (\bibinfo{year}{2003}).

\bibitem[{\citenamefont{Meier and {Zakharchenya (Eds.)}}(1984)}]{Meier:1984}
\bibinfo{author}{\bibfnamefont{F.}~\bibnamefont{Meier}} \bibnamefont{and}
  \bibinfo{author}{\bibfnamefont{B.~P.} \bibnamefont{{Zakharchenya (Eds.)}}},
  \emph{\bibinfo{title}{Optical Orientation}}
  (\bibinfo{publisher}{North-Holand, New York}, \bibinfo{year}{1984}).

\bibitem[{\citenamefont{Fiederling et~al.}(1999)\citenamefont{Fiederling,
  Kleim, Reuscher, Ossau, Schmidt, Waag, and Molenkamp}}]{Fiederling1999:N}
\bibinfo{author}{\bibfnamefont{R.}~\bibnamefont{Fiederling}},
  \bibinfo{author}{\bibfnamefont{M.}~\bibnamefont{Kleim}},
  \bibinfo{author}{\bibfnamefont{G.}~\bibnamefont{Reuscher}},
  \bibinfo{author}{\bibfnamefont{W.}~\bibnamefont{Ossau}},
  \bibinfo{author}{\bibfnamefont{G.}~\bibnamefont{Schmidt}},
  \bibinfo{author}{\bibfnamefont{A.}~\bibnamefont{Waag}}, \bibnamefont{and}
  \bibinfo{author}{\bibfnamefont{L.~W.} \bibnamefont{Molenkamp}},
  \bibinfo{journal}{{\sl Nature}} \textbf{\bibinfo{volume}{402}},
  \bibinfo{pages}{787} (\bibinfo{year}{1999}).

\bibitem[{\citenamefont{Jonker et~al.}(2000)\citenamefont{Jonker, Park,
  Bennett, Cheong, Kioseoglou, and Petrou}}]{Jonker2000:PRB}
\bibinfo{author}{\bibfnamefont{B.~T.} \bibnamefont{Jonker}},
  \bibinfo{author}{\bibfnamefont{Y.~D.} \bibnamefont{Park}},
  \bibinfo{author}{\bibfnamefont{B.~R.} \bibnamefont{Bennett}},
  \bibinfo{author}{\bibfnamefont{H.~D.} \bibnamefont{Cheong}},
  \bibinfo{author}{\bibfnamefont{G.}~\bibnamefont{Kioseoglou}},
  \bibnamefont{and} \bibinfo{author}{\bibfnamefont{A.}~\bibnamefont{Petrou}},
  \bibinfo{journal}{Phys. Rev. B} \textbf{\bibinfo{volume}{62}},
  \bibinfo{pages}{8180} (\bibinfo{year}{2000}).

\bibitem[{\citenamefont{Young et~al.}(2002)\citenamefont{Young,
  {Johnston-Halperin}, Awschalom, Ohno, and Ohno}}]{Young2002:APL}
\bibinfo{author}{\bibfnamefont{D.~K.} \bibnamefont{Young}},
  \bibinfo{author}{\bibfnamefont{E.}~\bibnamefont{{Johnston-Halperin}}},
  \bibinfo{author}{\bibfnamefont{D.~D.} \bibnamefont{Awschalom}},
  \bibinfo{author}{\bibfnamefont{Y.}~\bibnamefont{Ohno}}, \bibnamefont{and}
  \bibinfo{author}{\bibfnamefont{H.}~\bibnamefont{Ohno}},
  \bibinfo{journal}{Appl. Phys. Lett.} \textbf{\bibinfo{volume}{80}},
  \bibinfo{pages}{1598} (\bibinfo{year}{2002}).

\bibitem[{\citenamefont{Jiang et~al.}(2003)\citenamefont{Jiang, Wang, van
  Dijken, Shelby, Macfarlane, Solomon, Harris, and Parkin}}]{Jiang2003:PRL}
\bibinfo{author}{\bibfnamefont{X.}~\bibnamefont{Jiang}},
  \bibinfo{author}{\bibfnamefont{R.}~\bibnamefont{Wang}},
  \bibinfo{author}{\bibfnamefont{S.}~\bibnamefont{van Dijken}},
  \bibinfo{author}{\bibfnamefont{R.}~\bibnamefont{Shelby}},
  \bibinfo{author}{\bibfnamefont{R.}~\bibnamefont{Macfarlane}},
  \bibinfo{author}{\bibfnamefont{G.~S.} \bibnamefont{Solomon}},
  \bibinfo{author}{\bibfnamefont{J.}~\bibnamefont{Harris}}, \bibnamefont{and}
  \bibinfo{author}{\bibfnamefont{S.~S.~P.} \bibnamefont{Parkin}},
  \bibinfo{journal}{Phys. Rev. Lett.} \textbf{\bibinfo{volume}{90}},
  \bibinfo{pages}{256603} (\bibinfo{year}{2003}).

\bibitem[{\citenamefont{Yonezu}(2002)}]{Yonezu2002:SST}
\bibinfo{author}{\bibfnamefont{H.}~\bibnamefont{Yonezu}},
  \bibinfo{journal}{Semicond. Sci. Technol.} \textbf{\bibinfo{volume}{17}},
  \bibinfo{pages}{762} (\bibinfo{year}{2002}).

\bibitem[{\citenamefont{Vurgaftman et~al.}(2001)\citenamefont{Vurgaftman,
  Meyer, and {Ram-Mohan}}}]{Vurgaftman2001:JAP}
\bibinfo{author}{\bibfnamefont{I.}~\bibnamefont{Vurgaftman}},
  \bibinfo{author}{\bibfnamefont{J.~R.} \bibnamefont{Meyer}}, \bibnamefont{and}
  \bibinfo{author}{\bibfnamefont{L.~R.} \bibnamefont{{Ram-Mohan}}},
  \bibinfo{journal}{J. Appl. Phys.} \textbf{\bibinfo{volume}{89}},
  \bibinfo{pages}{5815} (\bibinfo{year}{2001}).

\bibitem[{\citenamefont{Aperathitis et~al.}(1996)\citenamefont{Aperathitis,
  Kayiambaki, Foukaraki, Halkias, Panayotatos, and
  Georgakilas}}]{Aperathitis1996:ASS}
\bibinfo{author}{\bibfnamefont{E.}~\bibnamefont{Aperathitis}},
  \bibinfo{author}{\bibfnamefont{M.}~\bibnamefont{Kayiambaki}},
  \bibinfo{author}{\bibfnamefont{V.}~\bibnamefont{Foukaraki}},
  \bibinfo{author}{\bibfnamefont{G.}~\bibnamefont{Halkias}},
  \bibinfo{author}{\bibfnamefont{P.}~\bibnamefont{Panayotatos}},
  \bibnamefont{and}
  \bibinfo{author}{\bibfnamefont{A.}~\bibnamefont{Georgakilas}},
  \bibinfo{journal}{Appl. Surf. Sci.} \textbf{\bibinfo{volume}{102}},
  \bibinfo{pages}{208} (\bibinfo{year}{1996}).

\bibitem[{\citenamefont{Taylor et~al.}(2001)\citenamefont{Taylor, Jesser,
  Benson, Martinka, Dinan, Bradshaw, {Lara-Taysing}, Leavitt, Simonis, Chang
  et~al.}}]{Taylor2001:JAP}
\bibinfo{author}{\bibfnamefont{P.~J.} \bibnamefont{Taylor}},
  \bibinfo{author}{\bibfnamefont{W.~A.} \bibnamefont{Jesser}},
  \bibinfo{author}{\bibfnamefont{J.~D.} \bibnamefont{Benson}},
  \bibinfo{author}{\bibfnamefont{M.}~\bibnamefont{Martinka}},
  \bibinfo{author}{\bibfnamefont{J.~H.} \bibnamefont{Dinan}},
  \bibinfo{author}{\bibfnamefont{J.}~\bibnamefont{Bradshaw}},
  \bibinfo{author}{\bibfnamefont{M.}~\bibnamefont{{Lara-Taysing}}},
  \bibinfo{author}{\bibfnamefont{R.~P.} \bibnamefont{Leavitt}},
  \bibinfo{author}{\bibfnamefont{G.}~\bibnamefont{Simonis}},
  \bibinfo{author}{\bibfnamefont{W.}~\bibnamefont{Chang}},
  \bibnamefont{et~al.}, \bibinfo{journal}{J. Appl. Phys.}
  \textbf{\bibinfo{volume}{89}}, \bibinfo{pages}{4365} (\bibinfo{year}{2001}).

\bibitem[{gas()}]{gasi}
\bibinfo{note}{Ga$_{1-x}$Mn$_x$As was already grown on Si~\cite{Zhao2002:JCG}.}

\bibitem[{\citenamefont{Ishida et~al.}(2003)\citenamefont{Ishida, Sarma,
  Okazaki, Hwang, Ott, Fujimori, Medvedkin, Ishibashi, and
  Sato}}]{Ishida2003:PRL}
\bibinfo{author}{\bibfnamefont{Y.}~\bibnamefont{Ishida}},
  \bibinfo{author}{\bibfnamefont{D.~D.} \bibnamefont{Sarma}},
  \bibinfo{author}{\bibfnamefont{K.}~\bibnamefont{Okazaki}},
  \bibinfo{author}{\bibfnamefont{J.~O. J.~I.} \bibnamefont{Hwang}},
  \bibinfo{author}{\bibfnamefont{H.}~\bibnamefont{Ott}},
  \bibinfo{author}{\bibfnamefont{A.}~\bibnamefont{Fujimori}},
  \bibinfo{author}{\bibfnamefont{G.~A.} \bibnamefont{Medvedkin}},
  \bibinfo{author}{\bibfnamefont{T.}~\bibnamefont{Ishibashi}},
  \bibnamefont{and} \bibinfo{author}{\bibfnamefont{K.}~\bibnamefont{Sato}},
  \bibinfo{journal}{Phys. Rev. Lett.} \textbf{\bibinfo{volume}{91}},
  \bibinfo{pages}{107202} (\bibinfo{year}{2003}).

\bibitem[{\citenamefont{Cho et~al.}(2002)\citenamefont{Cho, Choi, Cha, Hong,
  Kim, Zhao, Freeman, Ketterson, Kim, and Kim}}]{Cho2002:PRL}
\bibinfo{author}{\bibfnamefont{S.}~\bibnamefont{Cho}},
  \bibinfo{author}{\bibfnamefont{S.}~\bibnamefont{Choi}},
  \bibinfo{author}{\bibfnamefont{G.-B.} \bibnamefont{Cha}},
  \bibinfo{author}{\bibfnamefont{S.~C.} \bibnamefont{Hong}},
  \bibinfo{author}{\bibfnamefont{Y.}~\bibnamefont{Kim}},
  \bibinfo{author}{\bibfnamefont{Y.-J.} \bibnamefont{Zhao}},
  \bibinfo{author}{\bibfnamefont{A.~J.} \bibnamefont{Freeman}},
  \bibinfo{author}{\bibfnamefont{J.~B.} \bibnamefont{Ketterson}},
  \bibinfo{author}{\bibfnamefont{B.~J.} \bibnamefont{Kim}}, \bibnamefont{and}
  \bibinfo{author}{\bibfnamefont{Y.~C.} \bibnamefont{Kim}},
  \bibinfo{journal}{Phys. Rev. Lett.} \textbf{\bibinfo{volume}{88}},
  \bibinfo{pages}{257203} (\bibinfo{year}{2002}).

\bibitem[{\citenamefont{Erwin and {\v{Z}uti\'c}}(2004)}]{Erwin2004:NM}
\bibinfo{author}{\bibfnamefont{S.~C.} \bibnamefont{Erwin}} \bibnamefont{and}
  \bibinfo{author}{\bibfnamefont{I.}~\bibnamefont{{\v{Z}uti\'c}}},
  \bibinfo{journal}{Nature Mater.} \textbf{\bibinfo{volume}{3}},
  \bibinfo{pages}{410} (\bibinfo{year}{2004}).

\bibitem[{\citenamefont{{\v{Z}uti\'{c}}
  et~al.}(2002)\citenamefont{{\v{Z}uti\'{c}}, Fabian, and {Das
  Sarma}}}]{Zutic2002:PRL}
\bibinfo{author}{\bibfnamefont{I.}~\bibnamefont{{\v{Z}uti\'{c}}}},
  \bibinfo{author}{\bibfnamefont{J.}~\bibnamefont{Fabian}}, \bibnamefont{and}
  \bibinfo{author}{\bibfnamefont{S.}~\bibnamefont{{Das Sarma}}},
  \bibinfo{journal}{Phys. Rev. Lett.} \textbf{\bibinfo{volume}{88}},
  \bibinfo{pages}{066603} (\bibinfo{year}{2002}).

\bibitem[{\citenamefont{Marka et~al.}(2003)\citenamefont{Marka, Pasternak,
  Rashkeev, Jiang, Pantelides, Tolk, Roy, and Kozub}}]{Marka2003:PRB}
\bibinfo{author}{\bibfnamefont{Z.}~\bibnamefont{Marka}},
  \bibinfo{author}{\bibfnamefont{R.}~\bibnamefont{Pasternak}},
  \bibinfo{author}{\bibfnamefont{S.~N.} \bibnamefont{Rashkeev}},
  \bibinfo{author}{\bibfnamefont{Y.}~\bibnamefont{Jiang}},
  \bibinfo{author}{\bibfnamefont{S.~T.} \bibnamefont{Pantelides}},
  \bibinfo{author}{\bibfnamefont{N.~H.} \bibnamefont{Tolk}},
  \bibinfo{author}{\bibfnamefont{R.~K.} \bibnamefont{Roy}}, \bibnamefont{and}
  \bibinfo{author}{\bibfnamefont{J.}~\bibnamefont{Kozub}},
  \bibinfo{journal}{Phys. Rev. B} \textbf{\bibinfo{volume}{67}},
  \bibinfo{pages}{045302} (\bibinfo{year}{2003}).

\bibitem[{\citenamefont{Lebedeva and Kuivalainen}(2003)}]{Lebedeva2003:JAP}
\bibinfo{author}{\bibfnamefont{N.}~\bibnamefont{Lebedeva}} \bibnamefont{and}
  \bibinfo{author}{\bibfnamefont{P.}~\bibnamefont{Kuivalainen}},
  \bibinfo{journal}{J. Appl. Phys.} \textbf{\bibinfo{volume}{93}},
  \bibinfo{pages}{9845} (\bibinfo{year}{2003}).

\bibitem[{\citenamefont{Fabian et~al.}(2002)\citenamefont{Fabian,
  {\v{Z}uti\'{c}}, and {Das Sarma}}}]{Fabian2002:PRB}
\bibinfo{author}{\bibfnamefont{J.}~\bibnamefont{Fabian}},
  \bibinfo{author}{\bibfnamefont{I.}~\bibnamefont{{\v{Z}uti\'{c}}}},
  \bibnamefont{and} \bibinfo{author}{\bibfnamefont{S.}~\bibnamefont{{Das
  Sarma}}}, \bibinfo{journal}{Phys. Rev. B} \textbf{\bibinfo{volume}{66}},
  \bibinfo{pages}{165301} (\bibinfo{year}{2002}).

\bibitem[{\citenamefont{Shockley}(1950)}]{Shockley:1950}
\bibinfo{author}{\bibfnamefont{W.}~\bibnamefont{Shockley}},
  \emph{\bibinfo{title}{Electrons and Holes in Semiconductors}}
  (\bibinfo{publisher}{D. {Van Nostrand}, Princeton}, \bibinfo{year}{1950}).

\bibitem[{\citenamefont{Levinshtein et~al.}(1996)\citenamefont{Levinshtein,
  Rumyantsev, and {Eds.}}}]{Levinshtein:1996}
\bibinfo{author}{\bibfnamefont{M.}~\bibnamefont{Levinshtein}},
  \bibinfo{author}{\bibfnamefont{S.}~\bibnamefont{Rumyantsev}},
  \bibnamefont{and} \bibinfo{author}{\bibfnamefont{M.~S.}
  \bibnamefont{{Eds.}}}, \emph{\bibinfo{title}{Handbook Series on Semiconductor
  Parameters}} (\bibinfo{publisher}{World Scientific, Singapore},
  \bibinfo{year}{1996}).

\bibitem[{tem()}]{temp}
\bibinfo{note}{Even if, at present, such spin polarizations are difficult to
  achieve at room temperature, theoretical calculations \cite{Erwin2004:NM} and
  high spin-polarization measurements at low temperature
  \cite{Braden2003:PRL,Panguluri2004:Pb} suggest that similar values should be
  feasible in ferromagnetic semiconductors at $T \sim 100$ K.}

\bibitem[{\citenamefont{Jonker}(2004)}]{Jonker2004:PC}
\bibinfo{author}{\bibfnamefont{B.~T.} \bibnamefont{Jonker}},
  \bibinfo{journal}{{private communication}}  (\bibinfo{year}{2004}).

\bibitem[{\citenamefont{Thompson}(2004)}]{Thompson2004:P}
\bibinfo{author}{\bibfnamefont{S.~M.} \bibnamefont{Thompson}},
  \bibinfo{journal}{{private communication}}  (\bibinfo{year}{2004}).

\bibitem[{\citenamefont{Bratkovsky and Osipov}(2004)}]{Bratkovsky2004:JAP}
\bibinfo{author}{\bibfnamefont{A.~M.} \bibnamefont{Bratkovsky}}
  \bibnamefont{and} \bibinfo{author}{\bibfnamefont{V.~V.}
  \bibnamefont{Osipov}}, \bibinfo{journal}{J. Appl. Phys.}
  \textbf{\bibinfo{volume}{96}}, \bibinfo{pages}{4525} (\bibinfo{year}{2004}).

\bibitem[{\citenamefont{Saikin et~al.}(2003)\citenamefont{Saikin, Shen, Cheng,
  and Privman}}]{Saikin2003:JAP}
\bibinfo{author}{\bibfnamefont{S.}~\bibnamefont{Saikin}},
  \bibinfo{author}{\bibfnamefont{M.}~\bibnamefont{Shen}},
  \bibinfo{author}{\bibfnamefont{M.~C.} \bibnamefont{Cheng}}, \bibnamefont{and}
  \bibinfo{author}{\bibfnamefont{V.}~\bibnamefont{Privman}},
  \bibinfo{journal}{J. Appl. Phys.} \textbf{\bibinfo{volume}{94}},
  \bibinfo{pages}{1769} (\bibinfo{year}{2003}).

\bibitem[{\citenamefont{Zhao et~al.}(2002)\citenamefont{Zhao, Matsukara, Abe,
  Chiba, Ohno, Takamura, and Ohno}}]{Zhao2002:JCG}
\bibinfo{author}{\bibfnamefont{J.~H.} \bibnamefont{Zhao}},
  \bibinfo{author}{\bibfnamefont{F.}~\bibnamefont{Matsukara}},
  \bibinfo{author}{\bibfnamefont{E.}~\bibnamefont{Abe}},
  \bibinfo{author}{\bibfnamefont{D.}~\bibnamefont{Chiba}},
  \bibinfo{author}{\bibfnamefont{Y.}~\bibnamefont{Ohno}},
  \bibinfo{author}{\bibfnamefont{K.}~\bibnamefont{Takamura}}, \bibnamefont{and}
  \bibinfo{author}{\bibfnamefont{H.}~\bibnamefont{Ohno}}, \bibinfo{journal}{J.
  Cryst. Growth} \textbf{\bibinfo{volume}{237-239}}, \bibinfo{pages}{1349}
  (\bibinfo{year}{2002}).

\bibitem[{\citenamefont{Braden et~al.}(2003)\citenamefont{Braden, Parker,
  Xiong, Chun, and Samarth}}]{Braden2003:PRL}
\bibinfo{author}{\bibfnamefont{J.~G.} \bibnamefont{Braden}},
  \bibinfo{author}{\bibfnamefont{J.~S.} \bibnamefont{Parker}},
  \bibinfo{author}{\bibfnamefont{P.}~\bibnamefont{Xiong}},
  \bibinfo{author}{\bibfnamefont{S.~H.} \bibnamefont{Chun}}, \bibnamefont{and}
  \bibinfo{author}{\bibfnamefont{N.}~\bibnamefont{Samarth}},
  \bibinfo{journal}{Phys. Rev. Lett.} \textbf{\bibinfo{volume}{91}},
  \bibinfo{pages}{056602} (\bibinfo{year}{2003}).

\bibitem[{\citenamefont{Panguluri et~al.}(2004)\citenamefont{Panguluri, Ku,
  Wojtowicz, Liu, Furdyna, {Lyanda-Geller}, Samarth, and
  Nadgorny}}]{Panguluri2004:Pb}
\bibinfo{author}{\bibfnamefont{R.~P.} \bibnamefont{Panguluri}},
  \bibinfo{author}{\bibfnamefont{K.~C.} \bibnamefont{Ku}},
  \bibinfo{author}{\bibfnamefont{T.}~\bibnamefont{Wojtowicz}},
  \bibinfo{author}{\bibfnamefont{X.}~\bibnamefont{Liu}},
  \bibinfo{author}{\bibfnamefont{J.~K.} \bibnamefont{Furdyna}},
  \bibinfo{author}{\bibfnamefont{Y.~B.} \bibnamefont{{Lyanda-Geller}}},
  \bibinfo{author}{\bibfnamefont{N.}~\bibnamefont{Samarth}}, \bibnamefont{and}
  \bibinfo{author}{\bibfnamefont{B.}~\bibnamefont{Nadgorny}}
  (\bibinfo{year}{2004}), \bibinfo{note}{preprint}.

\end{thebibliography}
\end{document}